# Software Engineering Process and Methodology in Blockchain-Oriented Software Development: A Systematic Study

Md Jobair Hossain Faruk*, Santhiya Subramanian*, Hossain Shahriar†, Maria Valero†
Xia Li*, Masrura Tasnim‡
*Department of Software Engineering and Game Development, Kennesaw State University, USA
†Department of Information Technology, Kennesaw State University, USA
‡Institute for Cybersecurity Workforce Development, Kennesaw State University, USA

{mhossa21, ssubram6, mtasnim1}@students.kennesaw.edu & {hshahria, mvalero2, xli37}@kennesaw.edu

*Abstract*—Software Engineering is the process of systematic, disciplined, quantifiable approach that has significant impact on large-scale and complex software development. Scores of well-established software process models have long been adopted in the software development life cycle that pour stakeholders towards the completion of final software products development. Within the boundary of advanced technology, various emerging and futuristic technology is evolving that really need the attention of software engineering community whether the conventional software process techniques are capable to inherit the core fundamental into the futuristic software development. In this paper, we study the impact of existing software engineering processes and models including Agile, and DevOps in Blockchain-Oriented Software Engineering. We also examine the essentiality of adopting state-of-art concepts and evolving the current software engineering process for blockchain-oriented systems. We discuss the insight of software project management practices in BOS development. The findings of this study indicate that utilizing state-of-art techniques in software processes for futuristic technology would be challenging and promising research is needed extensively towards addressing and improving state-of-the-art software engineering processes and methodology for novel technologies.

## I. INTRODUCTION

According to IEEE Standard, Software Engineering (SE) is the application of a systematic, disciplined, quantifiable approach to the development, operation, and maintenance of software, and the study of these approaches is called the application of engineering to software [1]. The software engineering discipline contains a large number of domains that helps the practitioner to develop quality software application by providing systematic guidelines. Such guidelines or instructions in the form of programs that are being used to govern the computer system and process the hardware components are often called software engineering processes [2]. It is undeniable that following the fundamental of software engineering in general, software process or methodology, in particular, helps the stakeholders to elicit the requirements, design, develop, test, and maintenance of high-level, large, complex, and critical software applications with quality of final software products development. Misleading one of these processes may result in poor software product development.

Software development is a complex socio-technical activity where organizations need to establish and maintain robust software development processes for the quality of final software products [3]. A software process model provides the abstraction of the actual process or a simplified representation of a software process where every model represents a process from a specific perspective [4]. Debate on the effectiveness of various software development methodologies is a long been topic among the software engineering community where none of the approaches is perfectly suited to every type of software development setting. It is due to the vast involvement not only the technical knowledge and skills but also in many other factors, such as human, management, quality assessment, and cost in the software development process [5]. In order to enhance the efficiency and increase the productivity of software implementation, adopting software process standards are essential towards sustaining an optimal software process.

With the rapid growth of novel technology, software engineering process has evolved since the inception of the primary concept of this domain. In the last decades, scores of futuristic technology have emerged, artificial intelligence, blockchain, quantum computing for instance. Existing software engineering methodologies including Waterfall, Agile, and DevOps are proven effective approaches for the conventional software development process and it is time to evaluate the necessity of improvements of the aforementioned approaches towards providing futuristic technology-enabled effective process techniques. Waterfall Model for instance was introduced by Winston Royce in 1970 which is a linear-sequential methodology [6]. On the other hand, Agile process model was developed in 1990 and introduced in 2001 that consists of various methods, scrum for instance with the aim to manage the existing problem of any project. Besides, the

DevOps movement started to coalesce sometime between 2007 and 2008 by Patrick Debois that enables collaborative and multidisciplinary organizational effort to automate continuous rapid delivery software product features that shall increase efficiency through monitoring and measuring activities while guaranteeing their correctness and reliability [7], [8].

From various novel and futuristic technologies, blockchain technology has witnessed growing attention and a rapidly evolving domain that has been adopted in various fields including finance, healthcare, education, transportation, and voting [9]–[12]. Blockchain is a decentralized, distributed ledger of the network that stores digital records within enhanced trustworthiness environment with advanced smart contracts technology that provide faster transaction processing, transparent operation, user anonymity, auditability, and high scalability [13], [14]. In this paper, we intend to study the software engineering process for blockchain-based software systems that empowers the state-of-art software process and methodology. The primary contributions of the paper are as follows:

- We conduct a comprehensive review of software engineering process and its various methods including Agile and DevOps towards improving the existing software process by following state-of-art concepts.
- We provide an adequate overview of blockchain-oriented software engineering and how we can inherit software processes into BOSE.
- We discuss the challenges and limitations and provide future research direction on software processes in blockchain-oriented software development.

The rest of the paper is structured as follows: Section III provides a comprehensive review of Software Engineering Process and discusses the current trends. Section II provides research design and Methodology. While Section IV discusses on Blockchain-Oriented Software Engineering, presents an overview of Software Process Improvement (SPI) and Project Management for BOS development. Section V provide discussion and future research direction. Finally, Section VI concludes the paper.

## II. RESEARCH DESIGN & METHODOLOGY

### A. Prior Research

Mahdi Fahmideh *et al.* [14] present a paper on software engineering for blockchain-based software systems where the authors conducted a systematic literature review of the state-of-the-art in BBS engineering research from a software engineering perspective. The paper cover in-depth BBS engineering processes, models, development activities, principles, challenges, and techniques. The paper also discusses software engineering for the BBS system, a conceptual framework for BBS and introduces a BBS map, provides a survey on various topics including blockchain-based system engineering. The paper also discusses Agile-based BBS development, Model-driven BBS development, Architecture-based BBS development, Pattern-based BBS development, and Ontology-based BBS development which reflects the core principles of software engineering for blockchain-based software development.

Selina Demi *et al.* [15] provide an adequate overview of the software engineering applications by focusing on blockchain-oriented software engineering. The authors suggest the uses of software engineering in blockchain technology that has the potential in automating a variety of software engineering activities. The authors cover various software engineering principles including software requirements, software engineering process, software quality, software maintenance, configuration management, software engineering management and some other topics. The study indeed provides direction to the software engineering practitioners to focus on replacing centralized systems to decentralize frameworks.

Michele Marchesi *et al.* [16] introduce a blockchain-based software development process by adopting agile methodology. The framework intends to help software developers to gather the requirement, analyze, design, develop, test, and deploy Blockchain applications. The authors primarily focus on UML-based design that shall reflect the architecture of blockchain applications. The study provides a crucial direction in modeling software in the blockchain environment by utilizing UML diagrams including class, state, sequence, smart contract diagram for BOS development processes.

### B. Research Goals

This study aims to illustrate the current research landscape in software engineering process and methodology for blockchain-oriented Software Development. In accordance with the goal, we formulated the following research questions:

- RQ1: What is the software engineering process and how it can be intersected in blockchain-oriented software development?
- RQ2: What are the potential of software process improvement (SPI) and software project management in Blockchain-Oriented Software Engineering (BOSE)?
- RQ3: What are the key emerging and improvements that shall be carried out in BOSE methodology?

### C. Research Methodology

The systematic literature review [7], [17] has been conducted to find the current innovations that are either completely new or improved existing approaches for the study on the Software Engineering Process and Methodology in Blockchain-Oriented Software Development, depicted in Fig. 1. A "Search Process" was implemented to acquire research papers that address our topic of study [18]. Thus, specific search strings were applied during our analysis in scientific databases which contained the keywords, "Software Engineering Process", "Software Engineering Methodology", "Blockchain-Oriented Software Engineering", "Blockchain-Oriented Software Development", and "Software Process Improvement".

The scientific databases that were used for procuring these papers including: (i) Google Scholar (ii) IEEE Xplore, (iii) ScienceDirect, (iv) ACM, (v) Springer Link, and (vi) ResearchGate. We adopt a screening process to find the most relevant papers by studying the paper title followed by reading and understanding the abstract and conclusion from screened papers. An exclusion and inclusion process based on (i) duplicate papers (ii) full-text availability, and (iii) papers that are not related to the paper topic was conducted to prune off research papers that had aspects that were not related to our literature review as well as duplicates that appeared during the initial search. Table II displays the details of the inclusion and exclusion process.

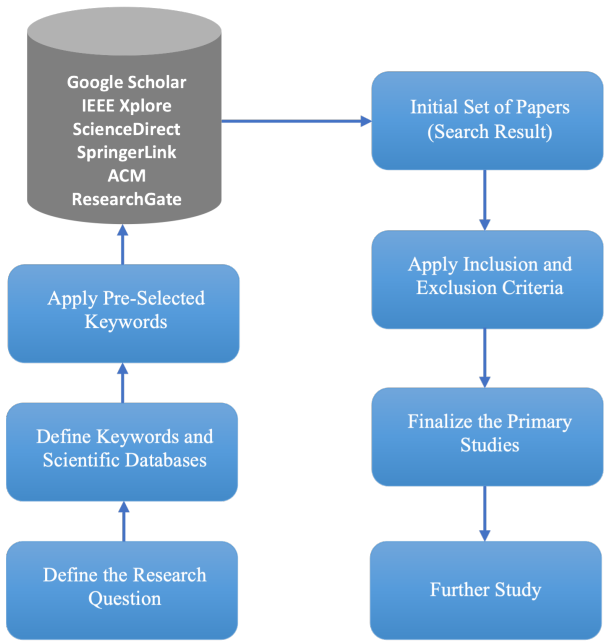

Fig. 1. Attrition of Systematic Literature through processing

TABLE I
GENERALIZED TABLE FOR SEARCH CRITERIA

| Scientific Database | Initial Keyword Search | Total Inclusion |
|---|---|---|
| Google Scholar | 138 | 9 |
| IEEE Xplore | 63 | 6 |
| ScienceDirect | 26 | 3 |
| Springer Link | 25 | 1 |
| ACM | 38 | 3 |
| ResearchGate | 64 | 2 |
| Total | 354 | 24 |

At first, the filtration procedure had a time constraint that allowed research papers published from the years 2016 to 2022. Furthermore, additional filters were placed in each database to narrow our search of relevant research materials. IEEE Xplore included Conferences and Journals while ScienceDirect required us to select Computer Science as the subject area and research articles for article type. We also included ACM scientific database where we included 3 papers while 1 and 2 papers are included from Springer Link and ResearchGate for this study. Total 354 research papers were found during the initial search but an in-depth screening process that accounted for the publication title, abstract, experimental results and conclusions shortened the list to 24 papers for our study.

TABLE II
INCLUSION AND EXCLUSION CRITERIA FOR THE PRIMARY STUDIES

| Condition (Inclusion) | Condition (Exclusion) |
|---|---|
| The study must be related to Software Engineering Process, Methodology, Blockchain-Oriented Software Development & SPI | Studies focusing on other topics than inclusion papers: blockchain in healthcare, business for instance |
| Papers are not duplicated in different databases | Similar papers in different databases |
| Studies are not duplicated in different scientific databases | Similar Studies in multiple scientific databases |
| Papers contain information related to software process and blockchain technology | Papers that do not cover expected domain |
| Peer-reviewed papers published in a conference proceeding or journal | Non-peer-reviewed papers |
| Studies that are available in the full format & in English | Studies are not available fully and Non-English studies |

### III. SOFTWARE ENGINEERING PROCESS

A software engineering process is a method that includes the whole run of exercises, from starting client beginning to software development and support [19]. Software Engineering Process also known as Software Development Life Cycle (SDLC) that involves strategies consisting of Analysis, Design, Coding, Testing, and Maintenance. A few distinctive prepare models exist and shift primarily within the frequency, application, and implementation of the methodologies mentioned above. Fig. 2 illustrates a basic flow of software engineering process initiated with problem statement to releasing the software product.

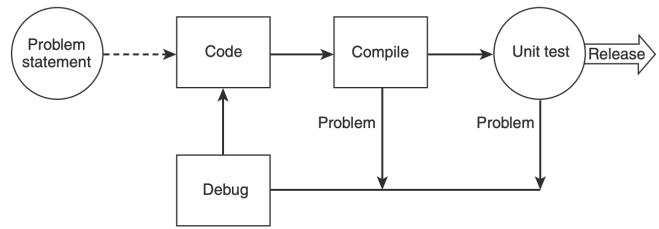

Fig. 2. illustrates a basic flow of software engineering process

Throughout the decades, various software development methodologies were proposed with the goal to provide support to develop good quality software. Software development methodology refers to the systematic processes that must be followed for producing the quality of software products by minimizing the time and cost-effective development. There are many software methodologies that have been established over time where the methodology helps the team to collaborate and communicate the information more efficiently [20]. Choosing the best software methodology for the project is crucial due to the pros and cons of each model. A methodology should be selected based on the project structure and nature since often choosing the wrong methodology leads the project failure.

*A. Waterfall*

For decades, the waterfall model has been used to describe a typical plan-driven approach to software development which is an approach to software development in the SDLC proces [21]s. A waterfall model is a linear and sequential approach to software development where the results of each phase are one or more documents that are signed-off and that the following phase should not start until the previous phase is completed. The waterfall model has different phases. Each phases output will be the input of the upcoming phases including feasibility study, requirement analysis and specification, design, coding, testing, and maintenance phase. While a waterfall approach may work well for simple, straightforward projects, it doesnât work well for complex projects.

*B. Agile*

Because of its versatility and responses to emerging technology, agile has become one of the most commonly used approaches to manage software projects [22]. Agile is not a single method, rather a collection of software development models and IT project procedures such as scrum, extreme programming practices, pair programming, and so forth. The agile manifesto conveys the values and principles of agile software development while comparing them to what is assumed to be the sole goal of a plan-driven methodology for software development [23]. Agile approaches foster a customer-centric culture within organizations that are made up of various approaches, all of which are founded on the principles of flexibility, transparency, quality, and continual progress. The fact that Agile project management focuses on both delivering quality and value to the customer and completing the project within the stipulated project limits is what makes it genuinely unique.

*C. DevOps*

DevOps is a combination of software development (Dev) and operations (Ops) and refers to a holistic approach focusing on cross-departmental integration and automation that enables organization to increase the delivery speed through process automation and reduce costs while continuing to meet control requirements [24], [25]. DevOps-oriented software development is a well-known trend adopted by many organizations dedicated cross-functional teams that work based on communication and collaboration. DevOps methodology provides software project teams working with small milestones addressing frequent and manageable delivery that allows significant efficiency in development, reduces the error in testing, and enables full automation from code to production [26]. Fig. 3 illustrates the architecture of DevOps that aims to bridge the gap between development and operational tasks.

Throughout the application lifecycle's plan, build, deliver, and operate phases, DevOps has a significant impact due to its interdependence nature while the phases are not divided into roles. Each position is involved in each phase to some level in a true DevOps culture while these approaches use technology to automate and optimize operations, it all starts with the culture within the organization and the people who contribute to it. Organizations that embrace a DevOps culture, on the other hand, can foster the growth of high-performing teams.

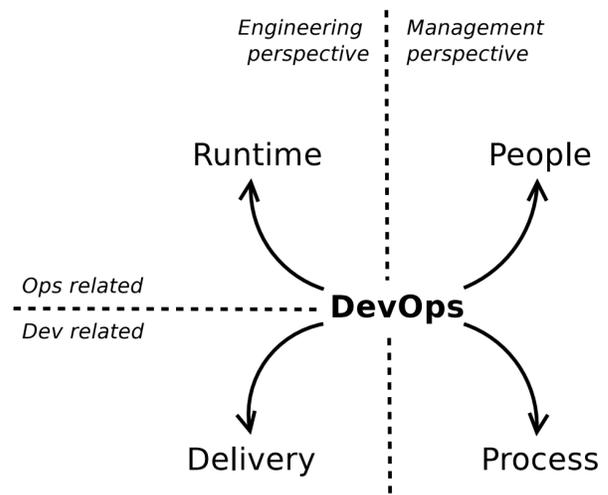

Fig. 3. depicts the overall conceptual map of DevOps [7]

Collaboration between teams, which originates with visibility, is one of the cornerstones of a healthy DevOps culture. By deploying applications in short cycles, DevOps teams stay agile. Even though development is incremental, shorter release cycles make planning and risk management easier, while substantially limiting the risks on system stability. One of the driving forces behind the DevOps culture is continuous integration and improvement. In reality, we can simply state that DevOps employs agile best practices to promote continuous integration via iterative development. The key benefit of continuous integration is that it makes it easier to spot flaws earlier in the development process rather than waiting until the end to integrate everything, which makes fixing errors more expensive. Continuous integration is essential for delivering valuable software quickly and frequently. It avoids waste by planning and executing value-adding activities in a logical order but requires agile concepts to revolutionize the way organizations build and deliver IT products and services. The keys to offering products and services to clients more frequently and pleasing them with value software at the press of a button are continuous integration and continuous delivery of valuable software.

## IV. BLOCKCHAIN-ORIENTED SOFTWARE DEVELOPMENT

Blockchain technology is an emerging and rapidly evolving domain that enables decentralized environment, stores digital records within enhanced trustworthiness blocks containing the unique security where each block connects with a hash of the previous block, forming in this way a chain of blocks

[27]–[29]. Towards enabling blockchain to inherit the concept of software engineering, a new domain has been emerged entitled "Blockchain-Oriented Software Engineering (BOSE)" [30], [31]. BOSE can define as a concept that adopts the principle of software engineering for providing an improved systematic, disciplined, quantifiable approach for blockchain-based software development.

### A. BOSE: Software Development

Adopting standard software engineering process and approaches for blockchain-oriented software introduced the novel field of blockchain-oriented software engineering [15]. BOSE is revolving at a staggering rate where the development process to gather the requirement, analyze, design, develop, test, and deploy blockchain applications is still in its infancy in consideration of the timeline [16], [32]. The goal here is to improve the current software process, techniques, and methodology to provide guidelines for systematically coordinating and controlling the tasks that must be performed in order to achieve the end product and the project objectives for blockchain software.

In traditional software development, ideal practices initiate with having a clear development process followed by design practices and implementation of the architecture, test, deployment, and security assessment practices can be continued after proper assessment and measuring the possible potential. In order to utilize the design and development process for blockchain applications and related smart contracts, following a process model is crucial. Scores of process models were introduced over the past few decades including Waterfall, Agile, DevOps, and RUP. Both Agile and DevOps provide an explicit development process including requirement elicitation, system design, specific notations, testing and security assessment. Since blockchain runs in a network of peer-to-peer nodes and smart contracts where both outputs and state must be the same in all nodes, the current process of methods needs to be improved based on blockchain specificities.

The primary complexity of blockchain-oriented software development is to find the suitable framework since there are various methods, permission, and permissionless blockchain in one side different framework including Ethereum and Hyperledger Fabric on another side. A group of researchers introduced a scrum-based iterative and incremental model for blockchain-oriented software development [33]. The model entitled as ABCDE integrate the blockchain components including smart contracts, libraries, data structures and in and out-of-chain components which all together constitute a complete decentralize software system. Fig. 4 shows the design method of the ABCDE model that focused on Ethereum blockchain and Solidity language. The framework initiates with the goal of the system, finding actors and user stories, designing the smart contract, coding, and testing both smart contract and app system, and finally integrating, testing, and deploying the decentralized system.

A group of researchers introduced a conceptual map of approaches of BBS software engineering depicted in Fig. 5

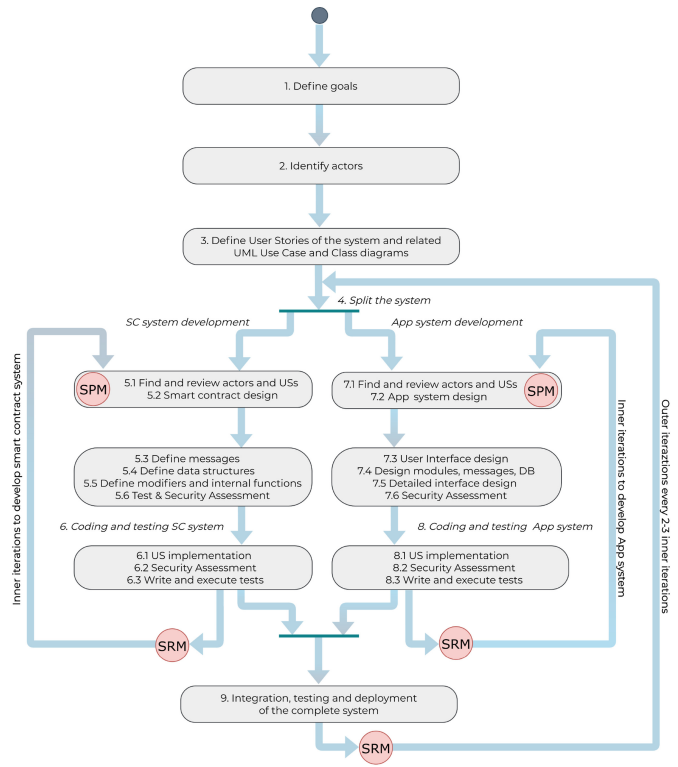

Fig. 4. ABCDE model for ethereum-based blockchain development [33]

focusing on five distinct themes including ontology-driven, patterns-based, architecture-based, model-driven, and agile-based development. It shall be noted that approaches are still in their infancy stage; however, adopting these techniques in the software development life cycle for blockchain-based applications may enable significant benefits. Such frameworks shall be helpful at the organizational level while many organizations prefer Hyperledger blockchain where we need to focus. It is obvious that Ethereum as a framework and Solidity as a programming language is one of the most popular blockchains; however, currently, Hyperledger blockchain is gaining attention due to its enterprise-based solution along with permission and consortium blockchain nature.

### B. DevOps in BOS Development

Recognizing the role of blockchain-oriented software development in software methodology including DevOps and Agile to manage changes in technology is crucial for the sustainability of software development in general [34]. DevOps software development model can provide a reliable and efficient environment for blockchain-oriented software development that shall increase organization's agility and accelerate disruptions during the development process. DevOps and blockchain-enabled workflows shall lead to faster software product development. Inheriting DevOps into blockchain-oriented software development requires a solid research foundation while there is no significant research so far has been found to address and resolve the challenges. A question arises of how DevOps can be adopted to the blockchain-based software process and approaches to project management? Besides, another research question comes whether blockchain can be implemented into DevOps to provide trust and security-enabled tools recorded within blockchain in a tamperproof way for use in audit and compliance? EY Global [24] introduces a Bi-modal DevOps implementation using blockchain while K. Vedagiri and R. Chandrasekaran [35] discuss the possible adoption of blockchain in DevOps tools Fig. 6 and operating DevOps within a transparent environment.

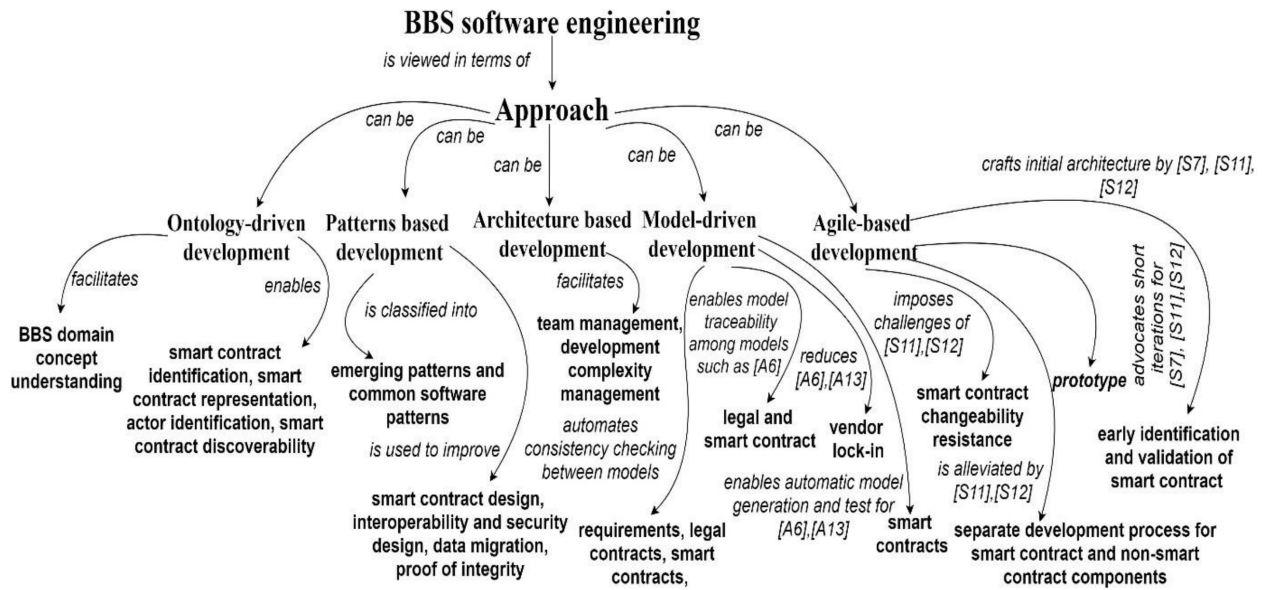

Fig. 5. illustrates a conceptual map (approach) of BBS software engineering [14]

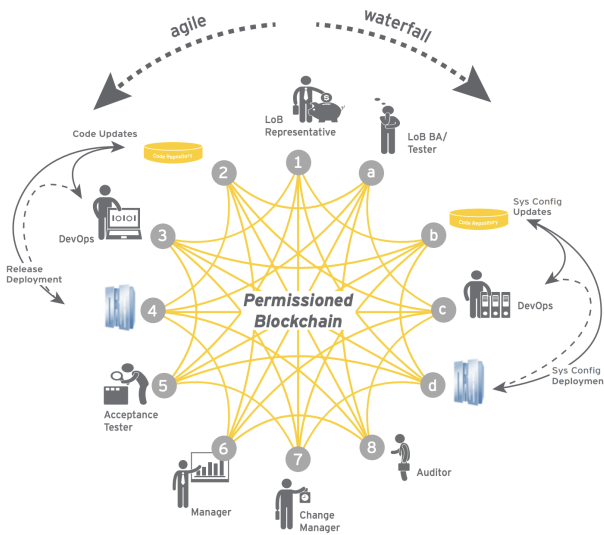

Fig. 6. Bi-modal DevOps implementation using blockchain [24]

### C. BOSE: Software Process Improvement (SPI)

Around the domain of BOSE, there is no significant study that implements blockchain-oriented SE applications in organizational settings due to the nascent phase of blockchain development along with the limited research, and the existence of unresolved technical challenges [15]. Effective research to identify the processes, methods, techniques, and tools for blockchain-oriented requirements engineering, software architecture, development, testing, and maintenance shall be addressed by the research community. Improving the existing software process for blockchain-oriented software development is a demanding topic that will help organizations to enhance the quality and efficiency of blockchain development by following the appropriate methodology throughout the blockchain-oriented software development life cycle (BO-SDLC).

Software Process Improvement (SPI) provides benefits to the organizations by enhancing the product quality, reducing development time, and cost of software projects. There are various existing methodologies and approaches that cause time and cost complexity in large-scale software development, particularly, when it comes to novel and emerging technology. Emerging blockchain-based software development needs focuses on SPI to overcome the challenges including reliance on a central body of standardization for certification, knowledge management, high cost, resource management and change in organizational culture. Usama Farooq *et al.* [36] named the domain as "Blockchain-Based Software Process Improvement (BBSPI)" and proposed a framework that has efficiency in SPI, reduces the time, cost, resources and helps to manage knowledge used to perform SPI. In accordance with the proposed framework, once a concerned organization or department reviews to improve the processes that involve software in general, the concerned entities shall place a request on the BBSPI portal for process improvements by providing their current organizational practices. The BBSPI will search for process engineers (peers) and select available peers to par-ticipate in the Software Process Improvement activity where a daisy-chain-styled network will be formed for improvement activity request.

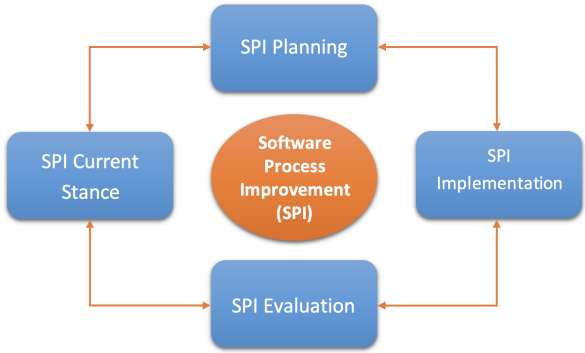

Fig. 7. depicts the software process improvement phases

Fig. 7 displays the common software process improvement phases by organizations that included SPI Planning, SPI Implementation, SPI Evaluation, and SPI Current Stance. SPI is important in Blockchain-oriented software development in order to improve projects cost by enhancing the process and avoiding issues, redundancies, and deficiencies. In future, the research community should focus on automated frameworks for BOSE-SPI for improved efficiency in BOS development.

*D. BOSE: Software Project Management*

Software Project Management (SPM) is to ensure to deliver a good quality product within the time limit, budget, and aligned with the business goals where stakeholders play a crucial role to choose the most appropriate process and methodologies to satisfy the desired products requirement. Project management includes team allocation, time management, budget, and maintenance along with four major phases including planning, organizing, monitoring, and adjusting. The decentralized blockchain technology that provides audibility and immutability has a different software management strategy in comparison with current conventional software development due to the differences from both technological and architectural aspects [37], [38]. Towards setting the baseline to allocate the project work, deadline, and budget based on the stakeholder request, the entire teams need to adopt a better model and framework. Managing the development of large-scale software systems is a challenge to all software project managers due to the complexity of software development and blockchain-enabled software development would be more challenging and this is where the research community should give attention. In this paper, we are yet to conduct high-level SPM research but in the future, we intend to conduct explicit research on Blockchain-oriented Software Project Management.

## V. DISCUSSION & NEW RESEARCH DIRECTION

Researchers in both software engineering and blockchain communities are working towards the establishment of newly emerging domain blockchain-oriented software engineering (BOSE), overcoming limitations and addressing the challenges. Based on our systematic study, we identify various research efforts by scores of researchers that cover areas including blockchain-based software process improvement (BB-SPI), project management, and blockchain in DevOps. Most of the frameworks are focused on permissionless blockchain due to the evolving nature of blockchain. Permissionless blockchain evolved early in 2009 while permissioned blockchain was introduced in 2015 by Hyperledger Foundation that is being adopted in various organizations for the unique architectural networks.

Blockchain-oriented software development still lacks a disciplined, organized, and mature development process while improvement is necessary in order to provide the best experience to blockchain developers. Future research should be devoted to blockchain-oriented software engineering and such efforts shall pave the way to address the limitation of the implementation of blockchain-enabled applications. In the future, our primary aim shall be conducting explicit research on terms including blockchain-oriented software engineering (BOSE) in general. The following research question should be addressed: what are the challenges in adopting the existing software process in Blockchain-oriented software engineering for requirements elicitation, design, development, and testing the decentralized software application?

Due to the novelty of blockchain-oriented software engineering, no significant effort was incorporated to solve the problems in traditional software process improvement and this paper highlighted various domains and intersected BOSE. Thus, the paper shall help the software engineering community and practitioners to practice from a new dimension by incorporating new blockchain-oriented techniques towards helping business analysts, architects, developers, programmers, and testers.

## VI. CONCLUSION

In this paper, we studied the software engineering process and its various methods including Agile and DevOps for better software process improvement (SPI) in blockchain-oriented software engineering (BOSE). We also provided an adequate overview of blockchain-oriented software development and how we can improve and inherit software processes for BOSE. Based on the research findings, various methods and frameworks have been introduced by researchers for BOS development. However, blockchain-oriented software engineering is still in its infancy and significant research is needs to be devoted to the comparably new blockchain framework including Hyperledger. Despite the challenges and limitations of BOS development, efforts in BOSE research are made to not only bridge this gap but also provide sound software engineering processes and practices.

## ACKNOWLEDGMENT

The work is partially supported by the U.S. National Science Foundation Awards 2100115, 1723578. Any opinions,

findings, and conclusions or recommendations expressed in this material are those of the authors and do not necessarily reflect the views of the National Science Foundation.